# Artificial Neural Networks Based Control and Analysis of BLDC Motors

Porselvi T, Sai Ganesh CS, and Aouthithiye Barathwaj SR Y

Porselvi T is with the Department of Electrical and Electronics Engineering, Sri Sairam Engineering College, Chennai, 600044 India (Phone: +91 94457 82722; e-mail: porselvi.eee@sairam.edu.in).

Sai Ganesh CS is with the Department of Electrical and Electronics Engineering, Sri Sairam Engineering College, Chennai, 600044 India (Phone: +91 8754444820; e-mail: saiganeshcs@ieee.org).

Aouthithiye Barathwaj SR Y is with the Department of Electrical and Electronics Engineering, Sri Sairam Engineering College, Chennai, 600044 India (Phone: +91 9840337075; e-mail: aouthithiyebarathwaj@ieee.org).

**ABSTRACT**

Artificial Neural Network (ANN) is the simple network that has input, output, and hidden layers with a set of nodes. Implementation of ANN algorithms in electrical, and electronics engineering always satisfies with the expected results as ANN handles binary data more accurately. Brushless Direct Current motor (BLDC motor) uses electronic closed-loop controllers to the switch DC current to the motor windings and produces the magnetic fields. The BLDC motor finds more applications because of its high speed, less maintenance and adequate torque capability. This motor is preferred to other motors due to its better performance and it is very easy to control its speed by Power Converters. This article presents a method of speed control of BLDC motor where speed is controlled by changing the DC input voltage of the bridge converter that feeds the motor winding. The control is done by using a PI based speed controller. The motor is modeled in the MATLAB/Simulink and the speed control is obtained with a PI controller. EMF signals, rotor speed, electromagnetic torque, Hall Effect signals, PWM and EMF signals simulations are obtained. The obtained data is fed into binary artificial neural networks and as a result, the ANN model predicts the corresponding parameters close to the simulation results. Both the mathematical based simulation and data based prediction gives satisfactory results.

## 1.  INTRODUCTION

The global climate change is one of the chronic disasters that include the global warming due to the emission of greenhouse gases by the technologies around us.  Several initiatives are taken especially in the field of transportation technology in order to reduce its effect towards global warming. Transportation alone contributes 29 percent in the global greenhouse gas emissions. The Electric and autonomous vehicles are impactful technologies in the global climate change and the conservation of fossil fuels. Brushless DC motors are the primary motors that are employed in electric vehicles. Electric vehicles generally employ more than two motors for propulsion. BLDC motor are basically a synchronous motor that uses direct current supply. In BLDC motors the mechanical commutator is replaced by electronic servo system making it reliable to detect the angle of rotor and to control the switches. And, electronic closed loop controllers are employed to switch Direct Current to motor windings to produce magnetic fields. The controller modifies the amplitude and phase of DC pulses to control the torque and speed of the motors. Bridge Converter is basically a DC-to-DC converter topology that employs four active switching components in a bridge configuration across a power transformer. A full bridge converter is a popular configuration that gives isolation as well as stepping up or down the input voltage. Reversing the polarity and providing multiple output voltages simultaneously are the additional functions provided with bridge converters. Speed control in BLDC motor has a significant role in the modern control systems. Open-loop and closed-loop methods are the two major types of control system. Dual closed-loop control is a popular term where the torque or current loop forms the inner control loop, and the voltage or speed loop forms the outer control loop. When the motor runs less than the rated speed, the input voltage of the motor is varied through Pulse Width Modulation strategy. When the motor runs beyond the rated speed, the flux is weakened as exciting current or auxiliary flux is advanced. Several methodologies are proposed to the speed-control control of BLDC motors. Commonly PID control is preferred as it is one of the most popular methods for many years and still employed in several applications. Moreover, PID controllers are employed extensively due to its robust property and more reliable. Generally, PID controllers satisfy the necessities of speed-regulation. Since BLDC motor is a non-linear system with multivariable, many challenges are to be considered for solving. At present, for speed control almost all BLDC motors employ PID controller for PWM.

Artificial intelligence is developing along with the evolution of computers and more applied in the simulation data in electrical and electronics engineering. Neural networks are a computational model that inherits the characteristics with the human brain that has several units running parallelly with no central control unit. Updating the weights is the key way the neural network to learn new information. In this paper, MATLAB/Simulink is used to simulate the speed control of BLDC motors. The simulation data is obtained and fed to artificial neural networks model. TensorFlow library is used for training the data in ANNs. Different ANN structures are proposed for different simulation data for optimal results.

## 2. TECHNICAL BACKGROUND

The Brushless Direct Current motors are applied everywhere from an electric vehicle to an electric fan. The higher starting torque is the main reason for its wider usage. The speed control of BLDC motors is becoming very important to suit its need for various applications. The use of fuzzy controllers to control the speed of the motors have been popularly in practice. The proper selection of fuzzy logic algorithm is also very important for efficient speed control[1-2]. There use of microcontroller hardware with fuzzy logic controller eliminates complex delay circuits. This reduces the overall cost of the system. In this method, the sensor-based fuzzy logic approach is not used to increase the robustness of the system [3]. In a paper by Suganthi P, the speed of the motor is using inverter via controlled voltage. This method dynamically controls the speed of the BLDC motor [4]. There have been a series of experiments being carried out for closed-loop speed control of BLDC motor. The experiments have shown better performance for very low speed to higher speeds by the rapid control prototyping method [5]. The performance of three-phase BLDC motor speed control using Fuzzy and PID controllers have been studied on various control system parameters like steady-state error, peak overshoot, rise, recovery and settling time [6]. Ibrahim in his paper compared Particle Swarm optimization and bacterial foraging techniques in order to determine the optimal parameter for PID controller of BLDC motor for speed control. His research output stated that the Particle Swarm optimization method can improve the dynamic performance of the system [7]. There have been a series of experiments conducted by Xiang wen, for speed control of BLDC motors. Through his experiments, results stated that the fuzzy PID controller with fuzzy control and PID small overshoot and small steady-state error has performed better in contrast to others [8]. The manual tuning of the PID controller is also being implemented in a three-phase BLDC motor with a six-step inverter. Without PID controller, the maximum overshoot observed is 25% and with PID controller it is nearer to zero [9]. Mahmud in his paper designed different control scheme with PID and fuzzy controller. The experimental results of his paper proved that the PID controller provides better performance out of the PI controller and fuzzy logic [10].In recent years, the use of ANN controllers increased exponentially. The ANN controller has a lesser settling time in contrast to other controllers. The response is nearly equal to the reference model. The model reference adaptive systems (MRAS) have also been implemented. The MRAS solves the problem of non-linearity and parameter variations. The ANN controller is robust, easy to implement and more effective than the PID and fuzzy controllers [11-13].

## 3. BLDC MOTORS

The BLDC motor has permanent magnets on its rotor and windings on its stator. Permanent magnets produce the rotor flux. The stator windings produce electromagnetic poles that attract the rotor and cause a rotation. The permanent magnets in BLDC motor rotate around an immobile armature, eradicating the challenges with linking current to the movable armature. The electronic controller in the BLDC motors, repeatedly switches the winding phases. Similar timed power distribution is produced by the controller when a solid-state circuit is used instead of a brush or commentator combination. The BLDC motor works in several phases, wherein the 3 phase motor is highly preferred and a popular system due to its high efficiency produces low torque and also accurate control.

## 4. ELECTRONIC CONTROLLERS

The proportional integral controller (PI) is the widely used control system than proportional integral derivative controller (PID). The main necessity for the controller is not only for speed control but also to reduce the error in the difference of actual and reference speed. The PI parameters mainly the gains i.e., Proportional, and integral gain affects the performance of the entire controller. Thus, parameter tuning is very important task and a difficult one. The PI controller is mainly used for the processes pertaining same input and disturbance and resulting in same output i.e., non-integrating. The Integral of Time-weighted Absolute Error (ITAE) and Internal Mode Control (IMC) methods are mainly used for tuning PI controllers. The current, voltage and speed measured are sent to the controller.

$$U(t) = K_p e(t) + K_i \int e(t) \qquad (1)$$

The working of the PI controller is based on the above equation. The error signal is calculated in the controller by taking difference of the actual and the reference values. In PI control the error is multiplied by proportionality and integral constant. The value we obtain will be in exponential order so in order to make it comparable it with other quantities it is passed on to a PWM signal converter. Then the PI controller controls the motor speed by changing the DC voltage fed in to motor winding through bridge converter as discussed above. A bridge converter is DC-DC converter which has better efficiency in contrast to a bridge rectifier. They enable to step-up and step-down inputs like a transformer and also offer isolation.

## 5. CONTROL PARAMETERS

### 5.1 Hall Effect

The commutation of a BLDC motor is electronically controlled unlike brushed motors. It is necessary to energize the stator windings as a sequence to rotate the motor. The position of the rotor is also necessary to know which winding needs to be energized. Hall Effect sensors that are fixed into the stator are employed to detect the position of the rotor. The sensors return high or low signal whenever the rotor magnetic poles advance towards the sensor thus indicating that the North or South pole is crossing the sensors. The sequence of commutation is thus determined based on the Hall sensors combination. According to the Hall Effect Theory, if a current carrying conductor is placed in a magnetic field, a transverse force is produced on the charge carriers that drive them to one side of the conductor. This is more observable in thin flat conductors. A voltage is produced between the two sides of the conductor when the magnetic influence gets balanced by the accumulation of charge at the sides of the conductor. This phenomenon is called the Hall effect.

## 5.2 Torque / Speed Characteristics

Two parameters for torque are used while defining a BLDC motor namely, Rated Torque (TR) and Peak Torque (TP). In the course of continuous usage, the motor can be loaded up till the Rated Torque. And as mentioned earlier, the torque remains constant until the speed reaches the Rated speed. The motor can be operated up to a maximum of 150% of the rated torque, but the torque takes a dip after a certain level. Certain applications require more torque than the Rated torque when it contains frequent reversals of rotation and frequent starts and stops with load on the motor. Such a situation arises mainly when the motor just starts and is also experienced during acceleration. Thus, some extra torque becomes necessary to control the effect due to inertia of load and the rotor. The motor torque can achieve a highest of up to peak torque, until it complies with the speed-torque curve.

## 5.3 Pulse Width Modulation

Pulse Width Modulation (PWM) is an efficient method for tuning the power delivered to the load. It is a technique of varying the speed in a smooth way without decreasing the starting torque. It also removes the harmonics. PWM is an efficient method for controlling the speed of a DC motor. A square wave with varying duty cycle is supplied to the motor. The duty cycle of PWM is dependent on the width of the pulse since the frequency is kept constant while the on-off time is varied. Thus, higher the power; higher the duty cycle in PWM. The switching logic is determined from the Hall signals to feed triggering pulses to the inverter switches. The actual and the reference signals are then compared and an error signal is provided to the PID controller, which then modifies the speed using the generated control signal accordingly.

## 6. ARTIFICIAL NEURAL NETWORKS

The Artificial Neural Network is a biologically inspired neural network that comes under the hood of Deep learning. The ANN consists of a connected set of points as a node called a neuron. The data processing and transformation are very similar to that of the data transfer and processing in the human brain i.e., by using synapses and neurons respectively. The

ANN structure mainly consists of an input layer, an output layer and more importantly a set of hidden layers. The input neuron receives input (data) and assigns weights to it and multiply them. Then they are added and passed on to the next neuron by applying some activation function. This process flows in a cyclic chain. This cyclic process enables the model to learn from the data enables perform complicated tasks. As discussed, the process starts by assigning weights to data moving between neurons. The main ability of the neural network to perform a task depends on the accurate adjustment of the weights. The weights are adjusted by the process of training the neural network with the training data. During training, the weights are changed for each iteration to make the model understand, more appropriately to identify the patterns in the given data. With the training, the trained neural network can predict new data fed into it. The accuracy depends on various parameters. Majority of the it depends on the size of the neural network, size of the dataset, activation function used and the type of the dataset.

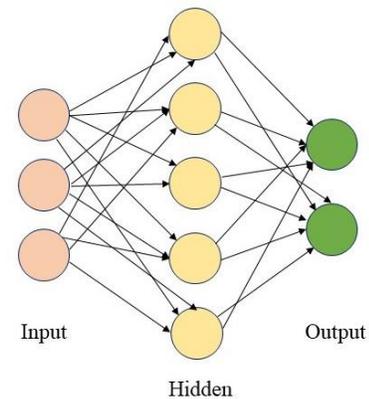

**Figure 1 Structure of a basic Artificial Neural Network**

The propagation of the neural network is based on two principles i.e., Forward propagation and Backward propagation. In forward propagation, the data is fed in. A sample is taken having features. These features determine the outcome. Then it performs to functions i.e., summation and activation. The assigned weights are summed up in the process. Then an activation function is applied and it is passed on to another hidden layer. The below equation where i is the number of inputs, W is the weights assigned and X are the feature(data).

$$Y = W_i X_i + b \qquad (2)$$

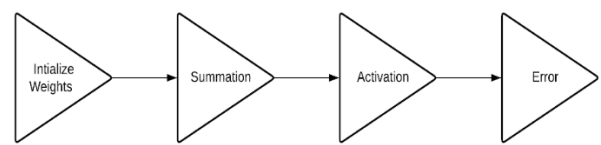

**Figure 2 Flow Diagram of Forward Propagation**

In Backward propagation, the propagation is mainly done to reduce the overall error. As in the name backward propagation, begins in error measurement and ending up in the modification of the weights to reduce the error measured through the activation function used and overall error measured. In Backward propagation, the chain rule is employed to find and rectify the error in weights responsible for the error as given below.

$$\frac{dE}{dW_i} = \frac{dE}{dA} \times \frac{dA}{dS} \times \frac{dS}{dW_i}$$

(3)

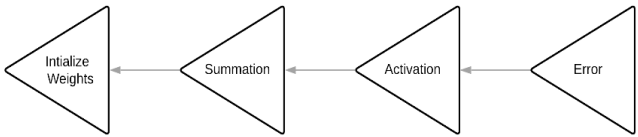

**Figure 3 Flow Diagram of Backward Propagation**

## 7. PROPOSED SYSTEM

The system proposed is the effective method of control of the Brushless Direct Current Motor (BLDC). The proposed method is based on the controlling of input DC Voltage of the bridge converter. The input DC voltage of the bridge converter is controlled as it feeds the motor windings which creates a magnetic field and in turn rotates the motor. The input of DC voltage of the bridge converted is controlled by using Proportional Integral controller (PI). The main gains i.e., Proportional, and Integral gain parameters are very important. They affect the entire performance of the PI controller. The PI controller is used here in order to obtain the same output i.e., non-integration output for the given input. The PI controller error calculation is by taking the difference of actual and reference values.

## 8. SIMULATION AND RESULTS

The system is simulated using MATLAB/Simulink. The circuit includes bridge converter, BLDC motor, PI controller and other necessary electrical components necessary to simulate the system. Figure 4 shows the simulation circuit of the system proposed.

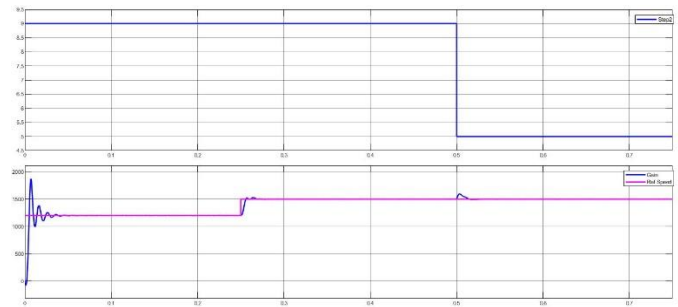

**Figure 5 Load Torque (Nm), Reference and Actual rotor speed (rpm)**

The Figure 5 has two graphs once is the time versus load torque and other is time versus actual and reference speed. The first one X axis represents time from 0-75 second interval. This X axis is common for both the graphs. The Y axes contains the load torque of the BLDC motor, after of about 50 seconds the load torque settles at 12 Newton-meter.

The other Y axis consists of two parameters i.e., reference and actual rotor speed. The actual rotor speed takes around 25 – 30 seconds to coincide with the reference speed.

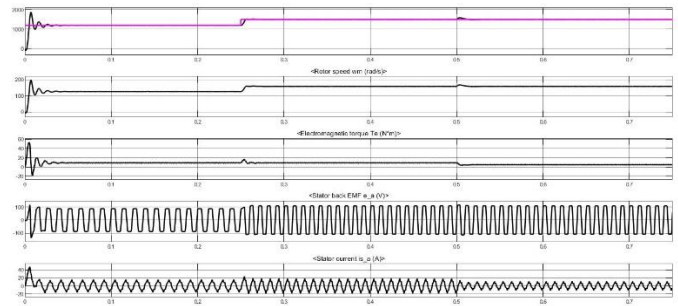

**Figure 6 Reference and actual rotor speed (rpm), rotor speed (rad/sec), electromagnetic torque (Nm), Stator back emf a-phase (V), Stator current a-phase (A)**

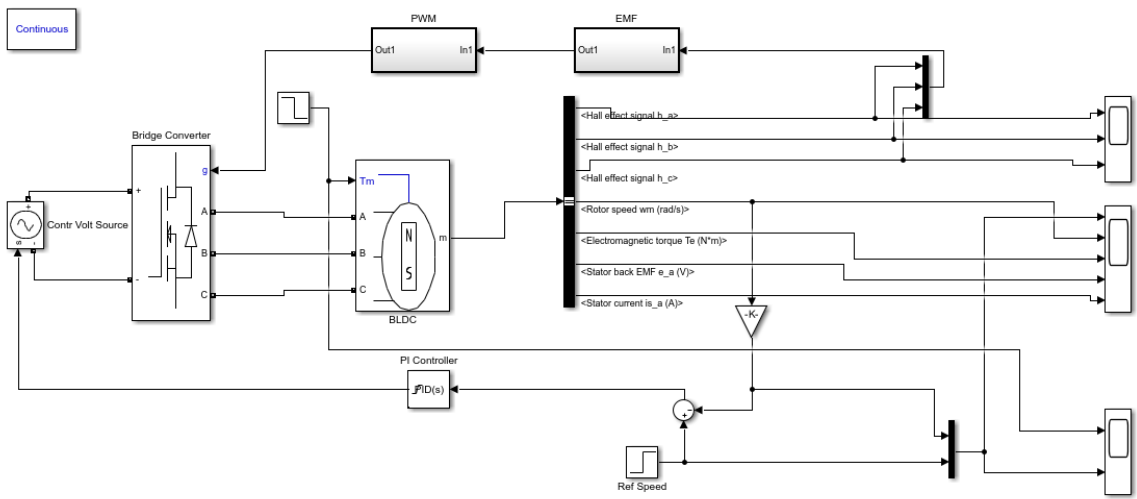

**Figure 4 Circuit Diagram of the Proposed System using Simulink**

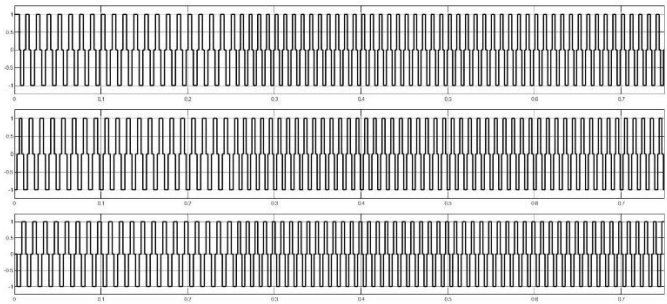

**Figure 7 EMF a, EMF b, EMF c (V)**

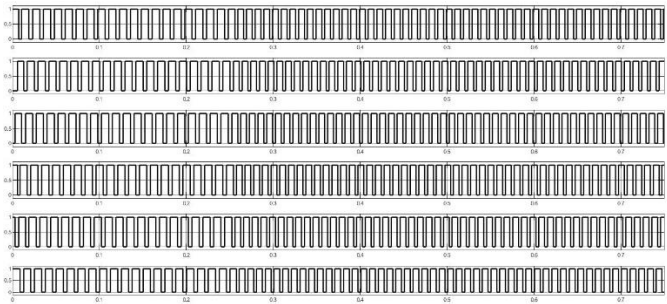

**Figure 8 PWM signals**

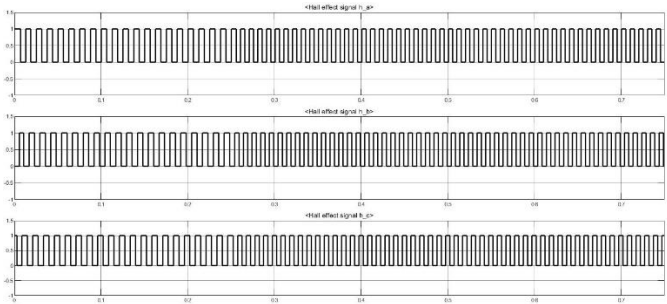

**Figure 9. Hall Effect signals**

The Figure 6, has 4 Y axes and a common X axis. The X axis is of a $0 - 75$ seconds time interval. The Y axes includes rotor speed, electromagnetic torque, stator back emf and stator current. The electromagnetic torque, stator back emf and rotor speed takes as low as 0.02 seconds to attain a fixed value. The stator current is in a constant variation over time to time.

The Figure 7, shows the emf of the BLDC motor. In this X is a taken as time and Y axes consists of EMF a, EMF b and EMF c. The EMF a, b, c constantly changes from time to time from -1 to 1 i.e., [-1,0,1].

The Figure 8, the time interval $0 - 75$ seconds is represented as X axis. The Y axes has 6 parameters of PWM signal i.e., PWM A, PWM B, PWM C, PWM D, PWM E and PWM F. The PWM signal is a binary output i.e., 0s and 1s.

The Figure 9, the time is taken along the X axis. The Hall Signal A, Hall Signal B and Hall C all along the Y axis. The Hall signal varies with the time between 0 and 1.

# 9. PREDICTION AND RESULTS

The data generated from the simulation of Brushless Direct Current motor with PID controller simulation are collected in excel sheet format and then converted into .csv format. The multiple parameters obtained from the simulation results are compared with each other and fed into an Artificial Neural Network (ANN). The experiment is proceeded by first selecting parameters to compare and then fed them into an ANN built for it.

## 9.1 Case 1: Actual Speed Prediction

The actual speed of the motor over the period of 75 seconds. i.e., from 0 to 75-second interval is taken to be predictive output. The Load toque, electromagnetic torque, and the reference speed for the same interval of 0 to 75 seconds are taken as the input. The ANN is consisting of seven layers and 5 nodes in each layer except in the input and the output layer. The SoftMax activation function is predominantly used here. Since we go over a series of values the SoftMax activation function tends to perform better in this series of prediction. The metrics accuracy, loss, Mean Squared Error (MSE) and Mean Accuracy Error (MAE) are obtained as shown in figure [10-12]. Here the training and test set is in the ratio of 80:20 and a hundred iteration have been run to get the specified output. From the graphical representation figure [11], we can interpret that the dataset which contained exponential inputs in the data of electromagnetic torque has created a surreal difference in the training accuracy and the validation accuracy. The difference in the accuracy of the model is around 0.

## 9.2 Case 2: Stator Current Prediction

The stator current of the motor over the period of 75 seconds. I.e., from 0 to 75-second interval is taken to be predictive output. The Stator back emf for the same interval of 0 to 75 seconds are taken as the input. The ANN is consisting of sixteen layers and 5 nodes in each layer except the input and the output layers. The SoftMax activation function is predominantly used here. Since we go over a series of values the SoftMax activation function tends to perform better in this series of prediction. The metrics accuracy, loss, Mean Squared Error(MSE) and Mean Accuracy Error(MAE) are obtained as shown in figure [13-15]. Here the training and test set is split in the ratio of 80:20 and a hundred iteration have been run to get the specified output. From the graphical representation figure [13-15], the difference in the accuracy of the model is around 0.0012 similar to case 1. It is due to the same reason for the presence of exponential value in the dataset for the data of the stator back emf and stator current data.

## 9.3 Case 3: EMF A, EMF B, EMF C Prediction using Hall Signals

The EMF A,B,C are taken for the same time period of 75 seconds are taken as predictive output quantities. The Hall signals A,B,C are taken as the input. The ANN is consisting of eight layers and 5 nodes in each layer excluding the input and output layers. The sigmoid activation function is employed here as the data we use here is a binary data. The sigmoid activation and higher accuracy. Hence, we employ sigmoid activation function. The metrics accuracy, loss, Mean Squared Error(MSE) and Mean Accuracy Error(MAE) are obtained as shown in figure [16-18]. Here the training and test set is split in the ratio of 80:20 and a hundred iteration have been run to get the specified output. We can observe from the graph obtained as shown in figure [16-18], that the loss is getting reduced over the iterations

### 9.4 Case 4: EMF A, EMF B, EMF C Prediction using PWM

The EMF A,B,C are taken for the same time period of 75 seconds are taken as predictive output quantities. The Pulse Width Modulation A,B,C,D,E,F are taken as the input. The Artificial Neural Networks (ANNs) consist of eight layers and 5 nodes in each layer excluding the input and output layers. The sigmoid activation function is employed here as the data we use here is a binary data. The sigmoid activation function is always preferred for binary data due to its simplicity and higher accuracy. Hence, we employ sigmoid activation function. The metrics accuracy, loss, Mean function is always preferred for binary data due to its simplicity. Accuracy Error(MAE) are obtained as shown in figure [19-21]. Here the training and test set is split in the ratio of 80:20 and a hundred iteration have been run to get the specified output. We can observe from the graph obtained as shown in figure [19-21], that the loss is getting reduced over the iterations similar to the case 3.

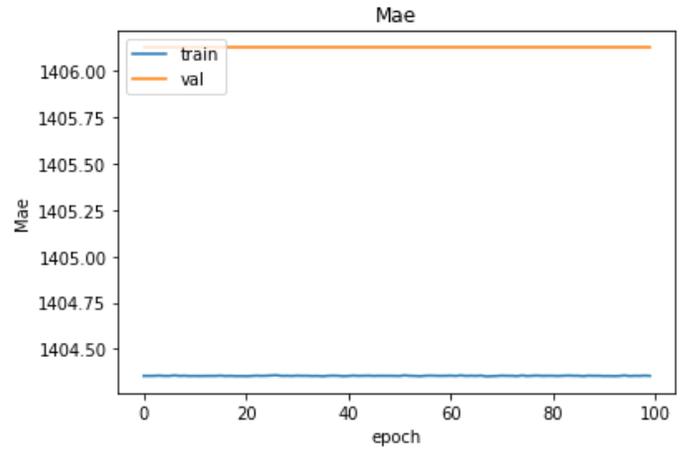

**Figure 12 Mean Average Error of Case 1**

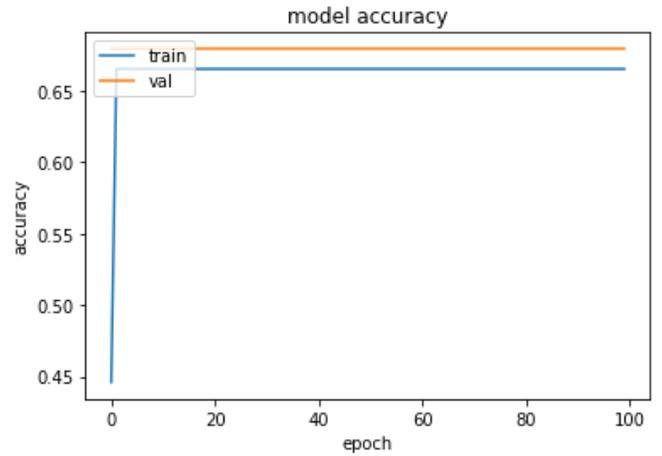

**Figure 13 Model accuracy of Case 2**

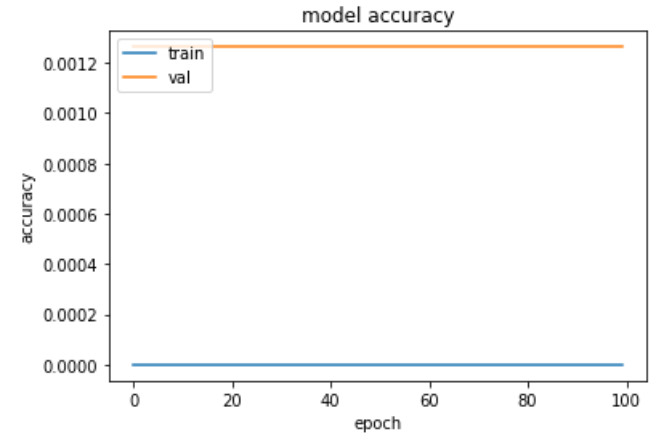

**Figure 10 Case 1 Model Accuracy**

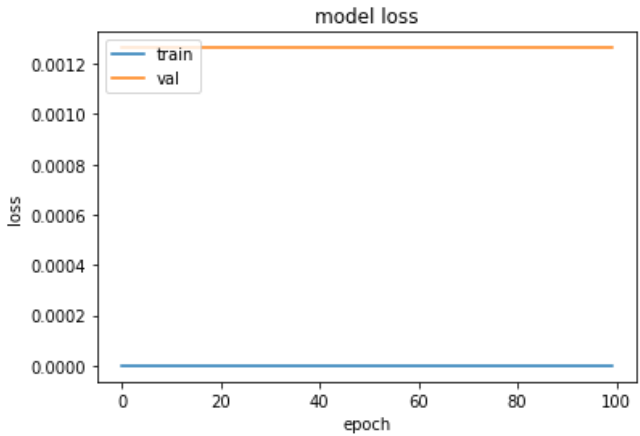

**Figure 11 Model Loss of Case 1**

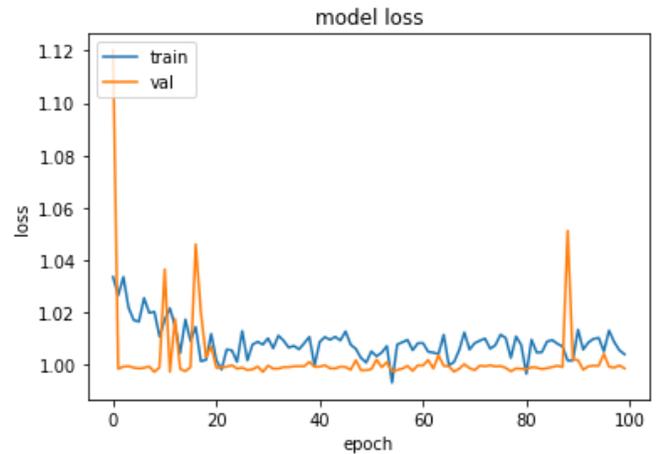

**Figure 14  Model Loss of Case 2**

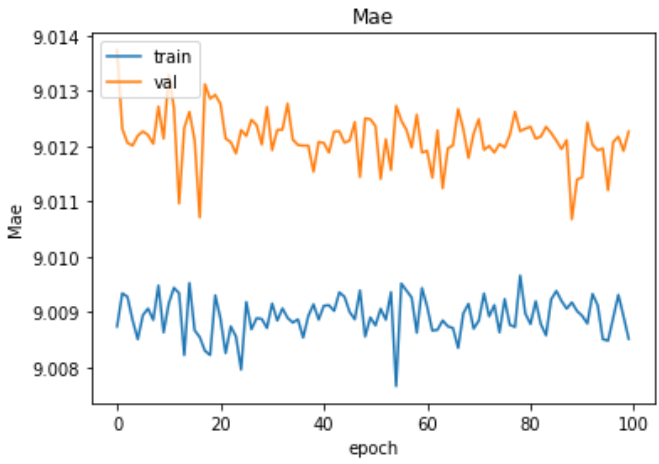

**Figure 15 Mean Average Error of Case 2**

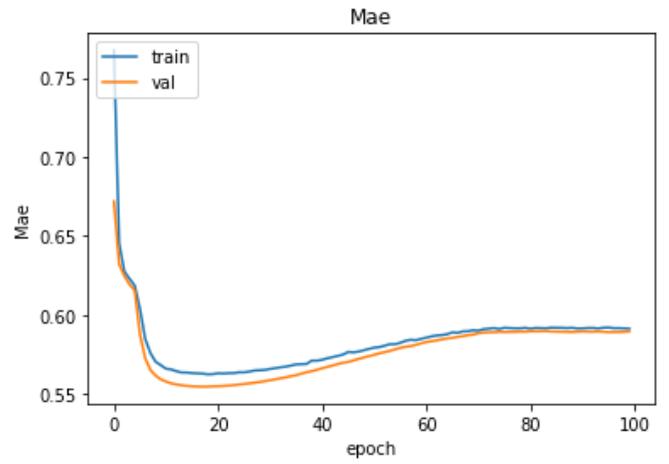

**Figure 18 Mean Average Error of Case 3**

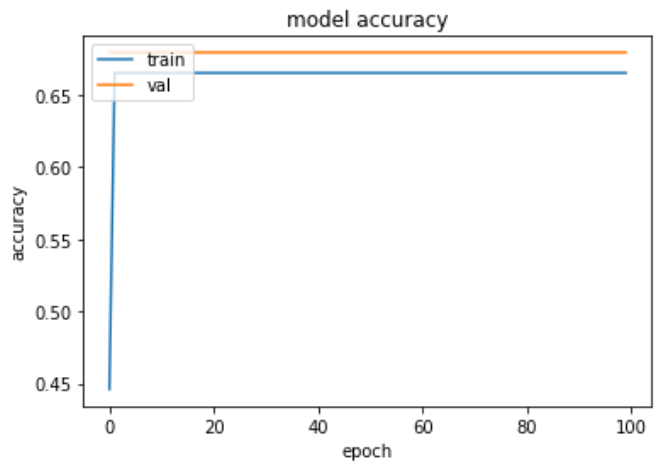

**Figure 16  Model Accuracy of Case 3**

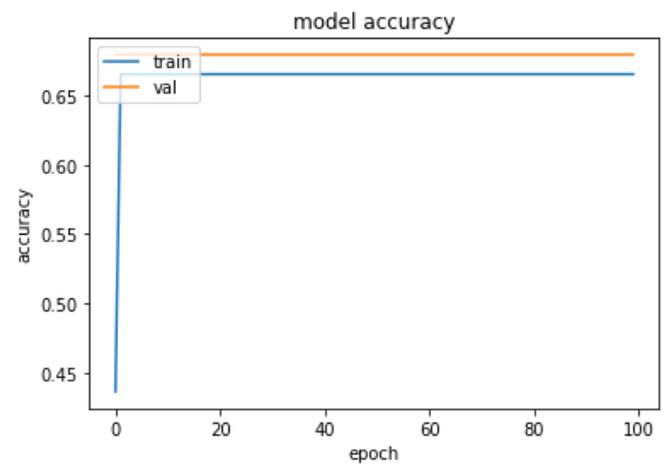

**Figure 19 Model Acurracy of Case 4**

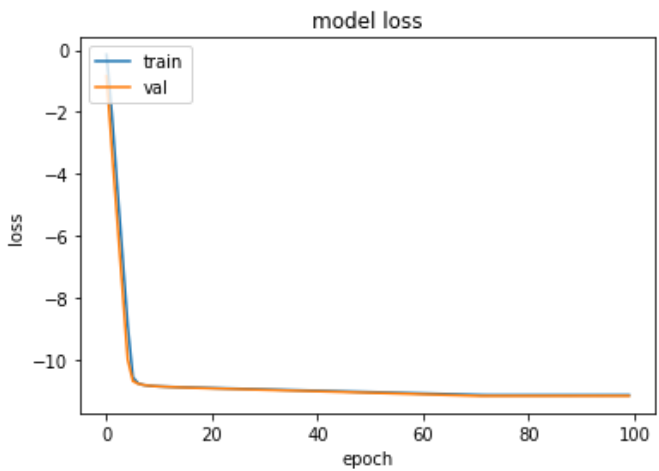

**Figure 17  Model Loss of Case 3**

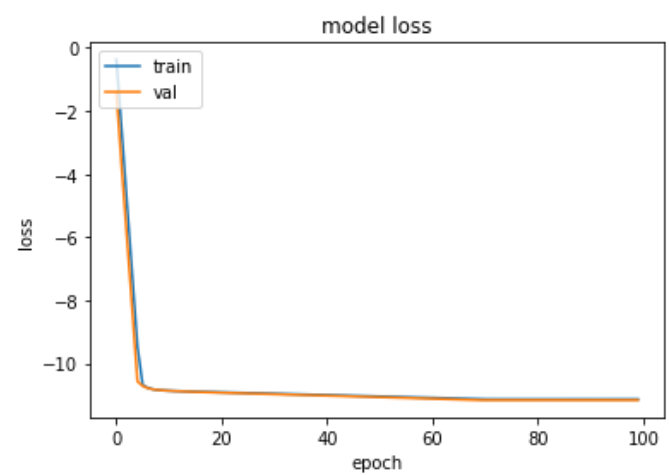

**Figure 20 Model Loss of Case 4**

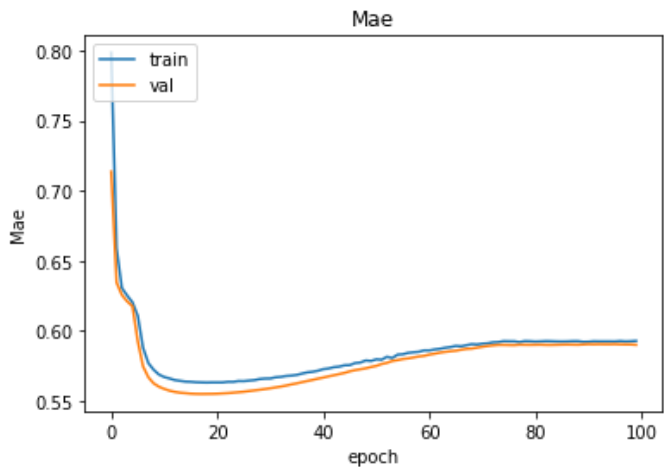

**Figure 21 Mean Average Error of Case 4**

## 10. Conclusion

In this paper, an effective method for speed control of BLDC motor is proposed. The proposed method is based on controlling the input of DC voltage of bridge converter which feeds the motor windings. The input is controlled using PI controller. For the proposed system the data is fed into to a binary Artificial Neural Network for various parameters obtained from the simulation. The simulation results and prediction result so obtained are synonymous. The activation function and neuron for each layer is altered and studied with respect to the input data.